# A MORPHOLOGY–COSMOLOGY CONNECTION FOR X–RAY CLUSTERS

August E. Evrard

*Physics Department, University of Michigan, Ann Arbor, MI 48109*

and

Joseph J. Mohr, Daniel G. Fabricant & Margaret J. Geller

*Harvard–Smithsonian Center for Astrophysics, Cambridge, MA 02138*

# A MORPHOLOGY–COSMOLOGY CONNECTION FOR X–RAY CLUSTERS

August E. Evrard

*Physics Department, University of Michigan, Ann Arbor, MI 48109*

and

Joseph J. Mohr, Daniel G. Fabricant & Margaret J. Geller

*Harvard–Smithsonian Center for Astrophysics, Cambridge, MA 02138*

## Abstract

We employ N–body/3D gas dynamic simulations of the formation of galaxy clusters to determine whether cluster X–ray morphologies can be used as cosmological constraints. Confirming the analytic expectations of Richstone, Loeb, & Turner, we demonstrate that cluster evolution is sensitive to the cosmological model in which the clusters form. We further show that evolutionary differences are echoed in the gross morphological features of the cluster X–ray emission.

We examine current–epoch X–ray images of models originating from the same initial density fields evolved in three different cosmologies: (i) an unbiased, low density universe with $\Omega_o = 0.2$; (ii) an unbiased universe dominated by vacuum energy with $\Omega_o = 0.2$ and $\lambda_o = 0.8$ and (iii) a biased Einstein–deSitter model ($\Omega = 1$, $\sigma_8 = 0.59$). Using measures of X–ray morphology such as the axial ratio and centroid shifting, we demonstrate that clusters evolved in the two low $\Omega_o$ models are much more regular, spherically symmetric, and centrally condensed than clusters evolved in the Einstein–deSitter model. This morphology–cosmology connection, along with the availability of a large body of cluster X–ray observations, makes cluster X–ray morphology both a powerful and a practical cosmological discriminant.

*Subject Headings:* cosmology: theory — galaxies: clustering — intergalactic medium X–rays: galaxies — hydrodynamics — methods: numerical

## 1. Introduction

Over the past fifteen years, observational studies have demonstrated that ∼40% of clusters exhibit evidence for multiple components (Geller & Beers 1982, Dressler & Shectman 1988, West & Bothun 1990, Jones & Forman 1990). Extensive theoretical studies have been carried out to explore the implications of the dynamical youth of some clusters and the details of cluster evolution (e.g., Cavaliere *et al.* 1986, Evrard 1990, Katz & White 1993). In particular, it has long been suspected that a connection exists between the evolution of clusters and the cosmological density parameter (Gunn & Gott 1972; Geller 1984); recent work by Richstone, Loeb & Turner (1992) has explored this relationship in detail. By modelling cluster growth as the collapse of Gaussian distributed, spherical density perturbations, they have shown that clusters form much earlier in low density ($\Omega_o \sim 0.2$) models than in $\Omega = 1$ models. This approach to cluster evolution necessarily omits a realistic treatment of the physical details of cluster formation, leaving questions regarding relaxation timescales, the effects of cluster growth on morphologies, and the effects of the surrounding large scale structure. We attempt to answer these questions with N–body/3$D$ gas dynamic simulations of galaxy clusters within their cosmic environment.

In this *Letter*, we demonstrate a *morphology–cosmology connection* for X-ray clusters and discuss its use as a cosmological diagnostic. The basic idea is that clusters in low $\Omega$ cosmologies are older and, hence, appear more relaxed and symmetric than their counterparts in an $\Omega = 1$ universe. Richstone *et al.* calculate the difference in median cluster ages between $\Omega_o = 0.2$ and $\Omega = 1$ to be $\sim 0.3$ $H_o^{-1}$. Because the crossing time for a 10 keV gas in the central 1 Mpc of a cluster is 0.6 billion years, this age difference corresponds to many sound crossing times. The observable effects of this age difference should be most apparent in the X-ray emission from the intracluster medium (ICM), because the isotropic pressure tensor of the gas drives it to lie on equipotential surfaces on a few sound crossing times, unless interupted by a merger event. To quantify this effect, we perform a suite of numerical cluster simulations (§2) using three viable cosmological backgrounds: (i) an unbiased, open universe $\Omega_o = 0.2$; (ii) an unbiased universe dominated by vacuum energy $\Omega_o = 0.2$, $\lambda_o = 0.8$ and (iii) a biased, Einstein–deSitter model $\Omega = 1$, $\sigma_8 = 0.59$. We analyse the cluster formation histories and X-ray morphologies in §3, and demonstrate that there are distinct differences in the cluster populations grown in high *versus* low $\Omega_o$ cosmologies. We discuss caveats in §4 and provide a summary in §5.

## 2. Numerical Models

In current theories, the nature of large-scale structure is controlled by a surprisingly small number of adjustable parameters. The present density parameter $\Omega_o \equiv 8\pi G \rho_o / 3H_o^2$ and dimensionless cosmological constant $\lambda_o \equiv \Lambda c^2 / 3H_o^2$ set the rate of change of scale factor with time, and thereby determine the growth rate of linear perturbations (Peebles 1980). The Hubble constant $H_o$ sets the normalization and, typically, one invokes a non–zero cosmological constant only to retain a flat metric if $\Omega < 1$, so that $\lambda_o = 1 - \Omega_o$. In addition, there are parameters required to describe the initial spectrum of linear perturbations. Except for models which invoke topological defects to generate perturbations (*e.g.*, strings or textures), the common assumption is that of a Gaussian random density field described by a power spectrum $P(k) \equiv |\delta_k|^2$, with $\delta_k$ the Fourier transform of the linear density perturbations. For reasonably smooth spectra, the spectrum in a finite range of wavelength centered on a scale of interest can be approximated as a power law with effective spectral index $n_{eff} \equiv d \ln P(k) / d \ln k$. For rich clusters of galaxies, the scale of interest is close to the fiducial normalization scale of $8\,h^{-1}$ Mpc ($h \equiv H_o/100$ km s$^{-1}$ Mpc$^{-1}$). Besides $n_{eff}$, the other parameter required to describe the initial density field is $\sigma_8$, the present, linear extrapolated variance $< (\delta\rho/\rho)^2 >^{1/2}$ within spheres of radius $8\,h^{-1}$ Mpc.

We generate a set of eight initial density fields from a standard cold dark matter (CDM) spectrum using a $32^3$ grid with Bertschinger's (1987) constrained technique. The constraint guarantees cluster formation in the simulation volume by requiring that the smoothed density field (smoothed with a Gaussian filter with $R_f = 0.2L$) evaluated at the box center be $2.5 - 5$ times the *rms* fluctuation on the filtered scale. We produce two density fields for each separate simulation volume; the simulation volume is a periodic cube of comoving length $L = 30$, $40$, $50$ or $60$ Mpc on a side (unless otherwise stated, all quoted lengths assume $h = 0.5$). Each initial realization is evolved in the three cosmological backgrounds listed above, making a total of 24 runs.

Density fields are sampled from an unbiased CDM spectrum with $\Omega h = 0.5$. For the $\Omega = 1$ models, we use the reduced normalization $\sigma_8 = 0.59$ ($\sigma_8 \equiv \langle (\delta\rho/\rho)^2 \rangle^{1/2}$ on the $8h^{-1}$ Mpc scale), equivalent to a bias parameter $b = 1.7$. This is the normalization inferred from the observed abundance of rich clusters (White, Efstathiou & Frenk 1993, Zabludoff & Geller 1993); an unbiased model favored by the COBE measurements produces too many rich clusters. We begin the $\Omega = 1$ simulations at an initial redshift $z_i = 9$ and reduce the initial density fields by a factor of $b(1 + z_i) = 17$. For each run, two sets of $32^3$ particles are displaced from a regular grid using the Zeldovich approximation applied to the generated density field. One particle species represents the baryons and the other the dark matter. The particle distributions are evolved in time using the combined N–body/$3D$ gas dynamic scheme P3MSPH (Evrard 1988). Gravity, $PdV$ work and shock heating are incorporated for the baryons. Radiative cooling is ignored. The minimum effective spatial resolution of the simulations is $\sim 0.005L$, which varies from 150 to 300 kpc for the box lengths employed. We approximate power on scales larger than the simulation volume by allowing for a DC component (typical initial amplitude $\sim 4\%$) in the density fluctuations; the volumes are then propagated in cosmologies with slightly higher background density.

In order to make direct comparisons of individual clusters evolved within different cosmological scenarios, we evolve each of the initial density fields in all three cosmological models. The initial redshifts for the low $\Omega_o$ runs are determined by requiring that the overall linear growth factor be the same as that in the Einstein–deSitter model. Requiring $\sigma_8 = 1$ (no biasing) at the present epoch in both low $\Omega_o$ models leads to an overall growth factor of 17 and starting redshifts of $z_i = 47.5$ for the $\Omega_o = 0.2$ model and $z_i = 23.0$ for the $\Omega_o = 0.2$, $\lambda_o = 0.8$ model. Reproducing the observed rich cluster abundance implies a normalization $\sigma_8 > 1$ (White, Efstathiou & Frenk 1993), but because we are not aware of any compelling mechanism for making galaxies *less* clustered than the mass, we have adopted a conservative "what you see is what you get" approach in the low density models. We assume a baryon fraction $\Omega_b = 0.1$ for *all* the models, allowing for some amount of collisionless matter (either dark matter or galaxies) in each cosmology.

Our approach has the advantage that the merger history of a particular initial density field is similar for all cosmological models. The merger *rate* depends on $\Omega$ and $\lambda$, but the proto–clusters or cluster 'building blocks' are (to zeroeth order) independent of the background cosmology. The price paid for this advantage is that the low $\Omega_o$ initial conditions are no longer consistent with a CDM spectrum. This is because features in the CDM power spectrum scale as the product $\Omega h$, which is fixed to the value 0.5. The appropriate value for the low $\Omega_o$ runs is $\Omega h \sim 0.2$. Using $\Omega h = 0.5$ causes the local spectral index $n_{eff}$ ($n_{eff} \equiv \mathrm{d} \ln P(k)/ \mathrm{d} \ln k$) to be slightly larger than it should be, roughly $-1$ rather than $-1.3$. This local spectral index controls the shape of the density distribution in collapsed halos, but the difference in halo profiles for the above spectral indices is very small (Crone, Evrard & Richstone 1993). We note that this range of spectral index is consistent with the power spectrum slope inferred from large–scale structure in the CfA redshift survey (Vogeley *et al.* 1992) and the 1.2 Jy and QDOT *IRAS* surveys (Fisher *et al.* 1993; Feldman, Kaiser & Peacock).

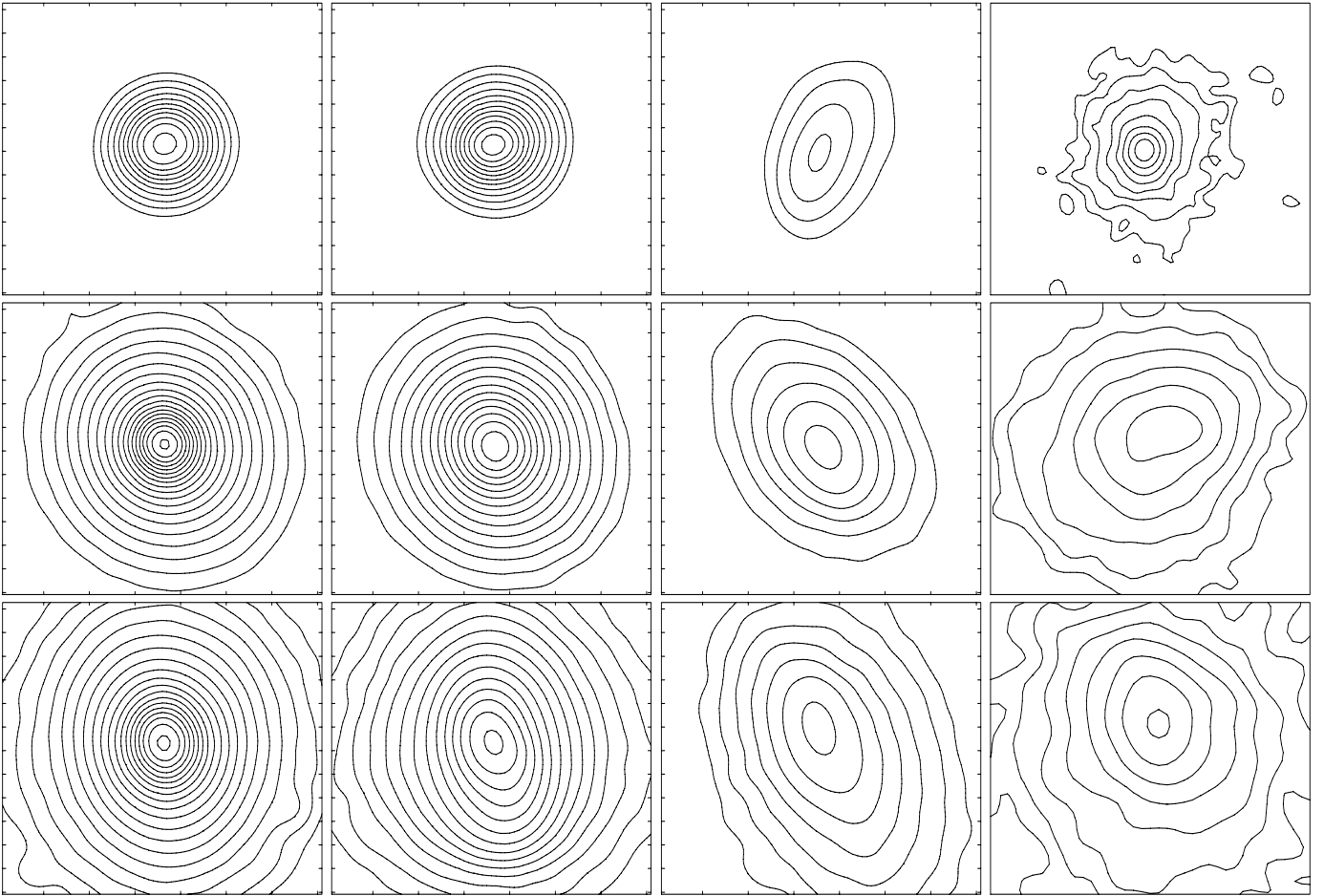

**Figure 1:** Contour maps of cluster X–ray emission: The first three columns show simulated clusters evolved in the following models: (i) $\Omega_o = 0.2$; (ii) $\Omega_o = 0.2$, $\lambda_o = 0.8$; and (iii) $\Omega = 1$, $\sigma_8 = 0.59$. The fourth column consists of *Einstein* Imaging Proportional Counter (IPC) observations of X–ray clusters. Each row in the first three columns corresponds to the same initial density field. The contours in all sixteen maps are spaced by factors of 1.8 in surface brightness from $3.5 \times 10^{-4}$ cts s$^{-1}$ arcmin$^{-2}$, and the spatial scale for every map is identical (the distance between tic–marks is 188 kpc). The IPC observations have been corrected for instrument response and Gaussian smoothed; from top to bottom, the clusters are: (i) A496, (ii) A399, (iii) A2256, (iv) and A401.

## 3. Comparison of X–ray Morphology

From the gas particle data for each simulation, we calculate X–ray maps (projected along three orthogonal axes) in the *Einstein* $0.5-3$ keV energy band following the procedure described in Evrard (1990). No Poisson noise or observational artifacts are added. Figure 1 contains contour maps of the X–ray emission from half of the simulated clusters. The first three images in each row represent the same initial density field evolved to the present epoch in the three different cosmological models. Clusters evolved in the low $\Omega_o$ universes are generally more spherical and centrally condensed than those evolved in the Einstein–deSitter model. This difference results from the accelerated evolutionary histories experienced by clusters in the low density universes.

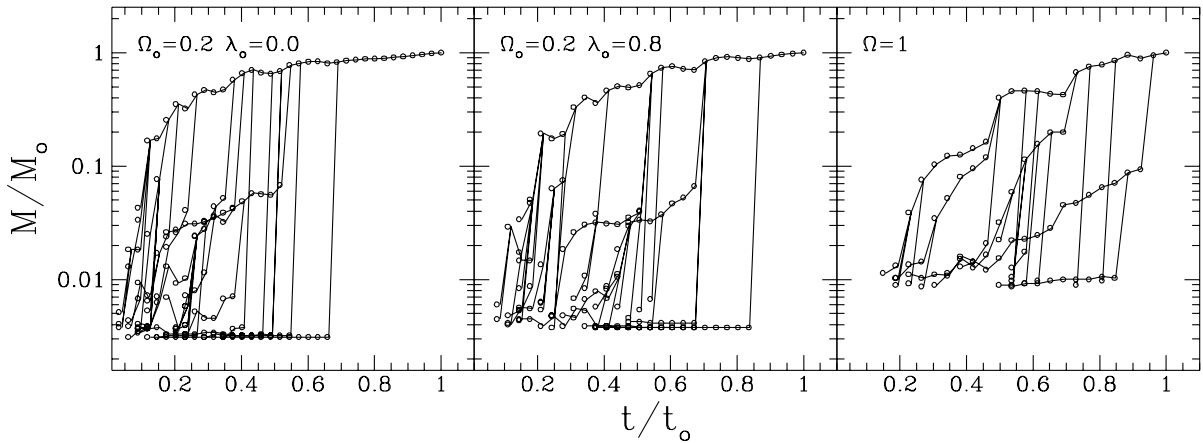

**Figure 2:** Merger hierarchy of the same initial density field (50a) evolved in the three different cosmological models. This plot follows the present–epoch cluster mass backward in time, identifying the protoclusters which eventually merge to form the final cluster. The vertical lines correspond to mergers of subclumps with the main cluster. The third row of Figure 1 shows contour maps of clusters formed from this initial density field.

We demonstrate this effect explicitly for the clusters in Figure 2 which were evolved in a 50 Mpc box. In this merger history diagram, we follow the mass associated with the final cluster backward in time, identifying the protoclusters which merge along the way to form the final cluster (the vertical lines correspond to mergers of subclumps with the main cluster; the algorithm is described fully in Evrard, Summers & Davis 1993). The cluster evolved in the $\Omega_o = 0.2$ model grows rapidly at early times (50% of its final mass is in place by $t/t_o = 0.3$) and relatively slowly at late times. In the Einstein–deSitter model, growth begins more slowly and continues into the present epoch; the cluster reaches 50% of its final mass at $t/t_o = 0.7$. Both the X–ray morphologies (Figure 1, row 2) and merger histories (Figure 2) indicate that the $\Omega_o = 0.2$, $\lambda_o = 0.8$ model is an intermediate case which is more similar to the $\Omega_o = 0.2$ model; the cluster contains 50% of its final mass at $t/t_o = 0.4 - 0.5$.

Also shown in Figure 1 are contour maps of *Einstein* X–ray images of four Abell clusters: A496, A399, A2256 and A401. Comparing these to simulated clusters indicates that the observations better match the clusters evolved in the Einstein–deSitter model. To compare clusters in these three models more quantitatively, we characterize the morphology in terms of two parameters: the axial ratio and the centroid shifting (Figure 3). Applying a method described in Mohr, Fabricant, & Geller (1993b), we allow for variations in centroid and axial ratio as a function of radius in the cluster by examining the images within annuli of constant width ($1'$) and varying radius (from $1'$ to $16'$). We determine the correct location of an annulus by minimizing the difference between the annulus center and the centroid of the photon distribution within the annulus. This approach avoids biasing the results by an *a priori* choice of image center. For each annulus, we then determine the image centroid and axial ratio. The axial ratio plotted in Figure 3 is the emission weighted mean axial ratio for the entire image, and $\Delta$, the centroid shift, is the emission weighted *rms* variation in centroid for the entire image. These measurements indicate that the two low $\Omega_o$ models produce clusters which are more regular and spherically symmetric than those evolved in the Einstein–deSitter model.

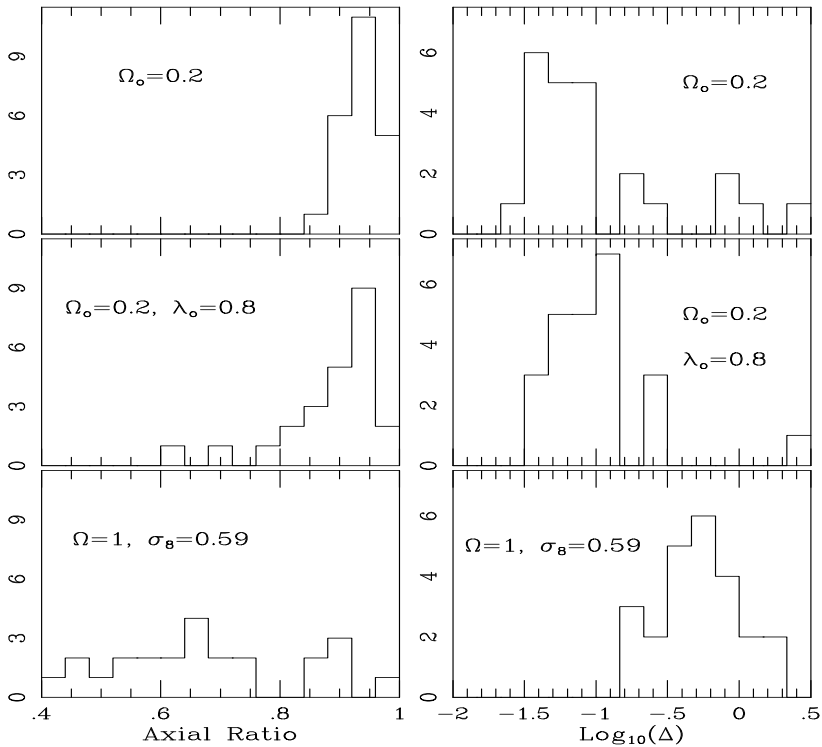

**Figure 3:** The distribution of axial ratio (left column) and centroid shift (right column) for all models (§3): Each row corresponds to a different cosmological model. $\Delta$, the centroid shift, is the emission weighted *rms* variation in centroid given in arc minutes ($1' = 70$ kpc); the axial ratio presented here is the emission weighted mean axial ratio. An inner region with a $6'$ radius is excluded in calculating the axial ratio and centroid shift of the $\Omega_o = 0.2$ models.

We are expanding on the work presented here by completing a systematic comparison of this battery of simulations with a representative ensemble of clusters observed by *Einstein*. By introducing the appropriate instrument reponse, accounting for X–ray point sources, and making the same measurements on simulated and observed clusters, we find that the clusters evolved in the Einstein–deSitter model most closely resemble observed clusters (Mohr *et al.* 1993a).

## 4. Discussion

Simulations, by nature, provide very specific information. One immediate and important question is how generic are the above results? What are the key controlling elements and how might they affect the results? We explore these issues here.

The rounder X–ray isophotes in the low $\Omega_o$ models result primarily from the scarcity of recent mergers; this allows the gas to relax onto isopotential surfaces. In addition, the isopotential contours are rounder in the low $\Omega_o$ clusters because the gas represents half of the mass. In contrast, the $\Omega\!=\!1$ clusters have 90% of their mass in the form of dark matter supported by an anisotropic velocity–dispersion tensor.

There are several potential concerns arising from the finite simulation volume. One is that the low $\Omega_o$ clusters may have had their accretion shut off artificially early. There are two lines of evidence discounting this idea. First, analytic models predict the early formation of clusters in low $\Omega_o$ universes (Richstone *et al.* 1992; Lacey & Cole 1993; Kauffmann & White 1993) and, second, the merger histories from these simulations agree well with merger histories calculated from an ensemble of clusters obtained from N–body simulations of larger volumes (Crone, Richstone, & Evrard 1993). This result is insensitive to the shape or normalization of the fluctuation spectrum (Lacey & Cole 1993) for reasonable definitions of the cluster population.

Another concern is that tidal effects generated on scales comparable to the simulation volume could distort the cluster gas, altering its X–ray morphology. However, extremely strong (and unrealistic) torques are required to affect gas in highly non–linear, X–ray emitting regions. For example, consider a point mass perturber of mass $M_p$ a distance $R$ from a cluster of mass $M_{cl}$. For the gravitational tides to have a fractional effect $f$ with respect to the cluster self–gravity on a test particle a distance $r$ from the cluster center would require

$$M_p \;>\; f \left(\frac{R}{r}\right)^3 M_{cl}.$$

For values $r\!=\!1$ Mpc and $R\!=\!L/4\!=\!7.5$ Mpc for our smallest box, a 10% effect on isophotes at 1 Mpc would require $M_p \gtrsim 40 M_{cl}$, which is unlikely.

Missing physics could also affect our results. It is difficult, however, to come up with mechanisms capable of strongly perturbing the low $\Omega_o$ gas morphology away from spherical symmmetry at the present epoch. Clusters evolved in the low $\Omega_o$ models have central densities sufficiently high that radiative cooling times are much less than the Hubble time. If radiative cooling (but no star formation and associated feedback) were turned on in the models, a large puddle of cold gas would collect in the cluster center, leaving a rarefied envelope of hot gas. There is no reason to suspect that the morphology of the hot envelope would be any less spherical than in the present models. Inclusion of galactic winds could add energy and metals to the gas but, again, generating an aspherical morphology would require (i) a *directed* energy input comparable to the total energy in the X–ray emitting gas ($\sim 10^{62}$ ergs) and (ii) that galaxy formation occur *after* cluster formation, so that the feedback occur very recently. These conditions are extremely unlikely. Simulations incorporating realistic feedback from galactic winds in an $\Omega = 1$ model show there is little effect on the gross morphological features of the X–ray emission at the present epoch (Metzler & Evrard 1993).

Finally, we have not explored a wide range of values of $\Omega_o$ and so we cannot explicitly state the sensitivity of cluster morphology to the exact value of $\Omega_o$. As $\Omega_o$ increases from 0.2 toward unity, the linear growth rates rapidly approach that for the $\Omega = 1$ model. In addition, adding a $\lambda_o$ component to flatten a model serves to make the growth curve more similar to the $\Omega = 1$ model. Our results, therefore, are better interpreted as an argument against $\Omega_o = 0.2$ than an argument for $\Omega = 1$. We hope to explore the $\Omega_o$ dependency of cluster morphology more extensively in future work.

## 5. Summary

We demonstrate a *morphology–cosmology connection* for X–ray clusters: evolutionary differences controlled by the underlying cosmological background are reflected in gross morphological features of the X–ray emission. This confirms the analytic expectation of Richstone, Loeb & Turner (1992), whose analysis was based on a spherical model of gravitational collapse. Using a combined N–body/3D gas dynamic method, we evolve eight different initial density fields to the present epoch in three different cosmological models: (i) $\Omega_o = 0.2$, $\sigma_8 = 1$; (ii) $\Omega_o = 0.2$, $\lambda_o = 0.8$, $\sigma_8 = 1$ and (iii) $\Omega = 1$, $\sigma_8 = 0.59$. We determine the axial ratio and centroid shifting of idealized 'observations' of these clusters and use these parameters to characterize the morphological differences in these models. Clusters evolved in the two low $\Omega_o$ models appear far more regular, spherically symmetric, and centrally condensed than clusters evolved in the Einstein–deSitter model. We are currently completing a direct comparison of simulated and real clusters using an ensemble of *Einstein* observations; results indicate that, of the three models tested, clusters evolved in the Einstein–deSitter model best match the complex nature of real cluster X–ray morphologies (Mohr *et al.* 1993a).

## References

Bertschinger, E. 1987, Ap. J., 323, L103.
Cavaliere, A., Santangelo, P., Tarquini, G. & Vittorio, N. 1986, Ap. J., 305, 651.
Crone, M., Evrard, A. E., & Richstone, D., 1993, in preparation.
Crone, M., Richstone, D. & Evrard, A. E., 1993, in preparation.
Dressler, A. & Shectman, S. A. 1988, A. J., 95, 985.
Evrard, A.E. 1988, MNRAS, 235, 911.
Evrard, A.E. 1990, Ap. J., 363, 349.
Evrard, A.E., Summers, F., & Davis, M. 1993, Ap. J., in press.
Feldman, H.A., Kaiser, N. & Peacock, J.A. 1993, Ap. J., submitted.
Fisher, K.B., Davis, M., Strauss, M.A., Yahil, A., & Huchra, J.P. 1993, Ap. J., 402, 42.
Forman, W.F. & Jones, C. J. 1990, in Clusters of Galaxies, eds. Oegerle, W. R., Fitchett, M. J., & Danly, L., (Cambridge: Cambridge University Press), 257.
Geller, M.J. 1984, Comments on Astrophysics & Space Science, 2, 47.
Geller, M.J., & Beers, T.C. 1982, P. A. S. P., 92, 421.
Gunn, J. & Gott, J.R. 1972, Ap. J., 176, 1.
Katz, N. & White, S.D.M. 1993, Ap. J.412, 455.
Kauffmann, G. & White, S.D.M., 1993, MNRAS, in press.
Lacey, C. & Cole, S. 1993, MNRAS, 262, 627.
Metzler, C.A. & Evrard, A.E. 1993, Ap. J., submitted.
Mohr, J. J., Evrard, A. E., Fabricant, D. G., & Geller, M. J. 1993a, in preparation.
Mohr, J. J., Fabricant, D. G., & Geller, M. J. 1993b, Ap. J., 413, in press.
Peebles, P.J.E. 1980, The Large-Scale Structure of the Universe, (Princeton : Princeton University Press).
Richstone, D., Loeb, A., & Turner, E. L. 1992, Ap. J., 393, 477.
Vogeley, M.S., Park, C., Geller, M.J. & Huchra, J.P. 1992, Ap. J., 395, L5.
West, M. J., & Bothun, G. D. 1990, Ap. J., 350, 36.
White, S.D.M., Efstathiou, G. & Frenk, C.S. 1993, MNRAS, 262, 1023.
Zabludoff, A. I. & Geller, M. J. 1993, A. J., submitted.